# Have underground radiation measurements refuted the Orch OR theory?[1]

## Kelvin J. McQueen

Philosophy Department, Chapman University

**Introduction**

In [1] it is claimed that, based on radiation emission measurements described in [2], a certain "variant" of the Orch OR theory has been refuted. I agree with this claim. However, the significance of this result for Orch OR *per se* is unclear. After all, the refuted "variant" was never advocated by anyone, and it contradicts the views of Hameroff and Penrose (hereafter: HP) who invented Orch OR [3].

My aim is to get clear on this situation. I argue that it is indeed reasonable to speak of "variants" here. Orch OR is not a complete model of reality but a work in progress. At its core, it claims that wavefunction collapse is a real physical event that has something to do with gravity ("OR") and that consciousness depends on orchestrated collapses in microtubules ("Orch"). There are many ways one could make these base ideas precise hence many "variants". Furthermore, the ways that HP aim to make these ideas precise are radical and incomplete. If they don't work out, Orch OR will need to fall back on another variant. Thus, I believe the significance of [1-2] for Orch OR is that it cuts out a small class of possible variants and leaves behind questions and challenges for the rest, including the variant preferred by HP.

**The refuted variant of OR**

In [2] it is claimed that radiation measurements have ruled out a version of OR (i.e. a certain model of wavefunction collapse), which the authors label "the natural parameter-free version of the Diósi-Penrose model". But this is unfortunate terminology. Following [4] it is reasonable to speak of "The Diósi-Penrose criterion for the rate of OR". For what Diósi and Penrose have in common is a formula for the lifetime of a superposition before it collapses. Following [4-5], the idea is that superpositions of spacetime curvature are unstable. The superposition lifetime is given by $t = \hbar/E_g$, where $E_g$ is the gravitational self-energy of the difference between the mass distributions belonging to the two states in the superposition. This formula has not been refuted. Nothing Penrose has put forward has been refuted. So, in what follows I refer to the refuted collapse model as the "parameter-free Diósi model".

When a charged particle gets accelerated, it produces a pulse of electromagnetic radiation. In most collapse models, a particle that undergoes collapse is likely to get accelerated. This is due to position/momentum uncertainty. A "perfect" collapse leaves the particle in a perfect position eigenstate, so that its momentum will be completely uncertain just after collapse. In the most well-studied collapse models (CSL, GRW), the solution is to replace "perfect" collapses with "imperfect" collapses, which leaves the particle in a "near" eigenstate of position, where "near" is defined by a new "localization resolution" length parameter, $\sigma$. The rate of collapse must also be controlled by a new parameter $\lambda$. Values for these parameters must be carefully chosen to be consistent with radiation (and other) measurements, see [6].

---

[1] This is an invited comment on [1] to be published in 2023 in Physics of Life Reviews.

The *parameter-free* Diósi model was proposed in the late 80's [7-8]. The role played by gravity meant that $\lambda$ could be replaced by $G/\hbar$ where G is Newton's constant. Meanwhile $\sigma$ could be replaced by a coarse-grained mass density operator with a spatial resolution $R_0$. This also plays a key role in determining $E_g$ and so the collapse rate. Diósi let $R_0$ be the nucleus length, making the model "parameter free". But this model was immediately criticized in 1990 for predicting "unacceptable detectable radioactivity" [9]. The proposed resolution was to let $R_0$ be an experimentally bounded free parameter. I'll call this resolution the *parameter-based* Diósi model. So, I think the real accomplishment of [2] is the bounds it places on this parameter. More recently, even stricter bounds have been placed by [10]. To be fair, [2] argues that parameter-based views are problematic because the parameter's value is unjustified as it is disconnected from the actual physics of the system. But I did not find this concern compelling. If $R_0$ is being determined experimentally, then there is a clear sense in which it is connected to the physics.

**The refuted variant of Orch OR**

Orch OR postulates that consciousness emerges from "orchestrated" collapses in microtubules. Recall: $E_g = \hbar/t$. HP calculate $E_g$ under certain assumptions, so as to count how many entangled tubulins are involved in these collapses. First: t = 25ms, because of 40hz oscillations of the neural correlates of consciousness. Second: the tubulin are not point-masses so HP represent their length with the carbon nucleus radius $a_c$ = 2.5 fermi. Finally: $a_c$ is used again as a measure of the distance of the separation of the two terms in the tubulin superposition. Now we can calculate $E_g$ and the number of tubulins involved. HP conclude that around $2 \times 10^{10}$ tubulins must be involved. There are roughly $10^9$ tubulins in one neuron. HP estimate that 0.001% of tubulins per neuron are involved and so conclude that around 20,000 neurons are involved in the superposition before collapse.

[1] argues that the Orch OR variant that uses the parameter-based Diósi model "is definitively ruled out for the case of atomic nuclei level of separation." It is ruled out because the superposition must instead involve $4 \times 10^{23}$ tubulins and $4 \times 10^{17}$ neurons – which is too many! (It is already controversial whether $2 \times 10^{10}$ tubulins could entangle without decoherence.) How did they get this result? The key assumption seems to be in the following passage:

> "Just as the [Diósi] collapse rate depends on $R_0$ so does the rate of spontaneous radiation. The larger $R_0$ the longer the collapse time and the lower the rate of spontaneous radiation emission. Conversely, if the resolution is fine, i.e., $R_0$ is chosen small like 2.5 Fermi then the predicted radiation becomes high enough to fall into the regime of experimental sensitivity."

So, if we increase the value of $R_0$ to avoid problematic radiation, we slow the collapse time. But the collapse time is *stipulated* to be t = 25ms. Therefore, increasing $R_0$ to an experimentally acceptable value while holding t = 25ms fixed entails that we need implausibly many tubulins and neurons in superposition in the brain. (It should be noted that in recent work [10] HP have moved on from the 25ms value to a range of values based on more up-to-date research on the physical correlates of consciousness.)

**Other Orch OR variants**

Above I explained how the Diósi model collapsed the two parameters of the CSL/GRW models into one parameter $R_0$. One question is whether this makes a difference to the argument of [1]. The authors do not address the issue, but I think should have. For their case against Orch OR is stronger the more variants they can rule out. But if their case only applies to the Diósi variant, then it is not a very pressing challenge. Thus, Orch OR variants that may be left open are variants based instead on CSL or GRW. Note that these will still fit the definition of OR that I gave above e.g. if the noise field in CSL depends on gravity somehow.

Although Penrose and Diósi agree on how to model the superposition lifetime, their collapse models differ dramatically. For Diósi collapse is gradual, for Penrose it's instantaneous. For Penrose, the radiation issue is dealt with by *retroactivity*: collapse induced accelerations are never large because collapses do not cause sudden localizations, instead they retroactively make it so that the particle was always relatively well-localized (see sec. 7 of [4]). This is an interesting (and mind-bending!) proposal. But unlike standard collapse models, it is lacking exact mathematical details. In [1] they argue that "to the extent that Penrose lacks a general mathematical model […] this can be regarded as a weak point of the Orch OR theory". But I think HP could flip this argument on its head and argue that the way standard collapse models (CSL, Diósi, GRW, etc) have used existing concepts to make their models precise can be regarded as a weak point, for those concepts have led to physical difficulties including spontaneous radiation and tension with relativity theory [4]. I would add that they've also led to conceptual difficulties, especially the tails problems, which I've elsewhere argued has not been adequately addressed [12]. It is therefore not unreasonable for HP to resist existing approaches to making collapse precise and hold out until we've developed better concepts.

A final variant worth mentioning is the fascinating recent suggestion of Diósi [13] in which energy conservation is recovered by postulating that collapse has a "frame-dragging" effect on the background spacetime. The upshot, then, is that there seem to be many variants of OR and Orch OR left open, with much work to do to render them precise enough for experimental tests. The research in [1-2] have made an important start by challenging some simple variants.